\documentclass[manuscript]{aastex}
\usepackage{txfonts}
\usepackage{graphicx}

\shorttitle{Molecular outflow and inflow in the Orion KL region}
\shortauthors{Wu et al.}

\begin{document}

\title{A study of dynamical processes in the Orion KL region using ALMA-- Probing molecular outflow and inflow}

\author{Yuefang Wu\altaffilmark{1}, Tie Liu\altaffilmark{2}, Sheng-Li Qin\altaffilmark{3}}

\altaffiltext{1}{Department of Astronomy, Peking University,
100871,Beijing China; ywu@pku.edu.cn }\altaffiltext{2}{Korea Astronomy and Space Science Institute 776, Daedeokdae-
ro, Yuseong-gu, Daejeon, Republic of Korea 305-348} \altaffiltext{3}{School of
Physical Science and Technology, Yunnan University,
Kunming,650091,China }

\begin{abstract}
This work reports a high spatial resolution observations toward
Orion KL region with high critical density lines of CH$_{3}$CN
(12$_{4}$-11$_{4}$) and CH$_{3}$OH (8$_{-1, 8}$-7$_{0, 7}$) as
well as continuum at $\sim$1.3 mm band. The observations were made
using the Atacama Large Millimeter/Submillimeter Array with a
spatial resolution of $\sim$1.5$^{\prime\prime}$ and sensitives
about 0.07 K and $\sim$0.18 K for continuum and line,
respectively. The observational results showed that the gas in the
Orion KL region consists of jet-propelled cores at the ridge and
dense cores at east and south of the region, shaped like a wedge
ring. The outflow has multiple lobes, which may originate from an
explosive ejection and is not driven by young stellar objects.
Four infrared bubbles were found in the Spitzer/IRAC emissions.
These bubbles, the distributions of the previously found H$_2$
jets, the young stellar objects and molecular gas suggested that
BN is the explosive center. The burst time was estimated to be
$\leq$ 1300 years. In the mean time, signatures of gravitational collapse toward Source I
and hot core were detected with material infall velocities of 1.5
km~s$^{-1}$ and $\sim$ 0.6 km~s$^{-1}$, corresponding to mass
accretion rates of 1.2$\times$10$^{-3}$M$_{\sun}$/Yr and
8.0$\times$10$^{-5}$M$_{\sun}$/Yr, respectively.  These
observations may support that high-mass stars form via accretion
model, like their low-mass counterparts.

\end{abstract}


\section{Introduction}
The dynamical processes in massive star formation region are still
not well understood. Molecular outflow and inflow are critical for
testing if high mass stars form in the same way as their low mass
counterparts. The Orion BN/KL is a massive star formation region
closest to us. The Orion KL region consists of different
spatial components including the "ridge" and "hot core".
The velocity ranges of these components are from 2.5 to 9
km~s$^{-1}$  \citep [and the references therein]{wri96}. Every
component harbors high-mass young stellar objects (YSOs)
identified with near and middle infrared emissions. And many of
the YSOs have subcomponents revealed by high
spatial resolution observations \citep [and the references
therein]{shu04}.

The first high velocity molecular outflow was identified and
discovered by use of CO J=1-0 line in Orion KL region
\citep{kwan76}. The outflow attracted extraordinary attention and
was further imaged by dense molecular tracers such as HCO$^{+}$
J=1-0 and SO (3$_2$-2$_1$) \citep [and the references
therein]{sch95,Fri84}. Besides the high velocity phenomena
detected in mm line emissions, finger-like H$_2$ filaments were
discovered in BN/KL region by near infrared observations also
\citep{tay84}. Line emissions of atoms, ions and high energy level
CO were detected in the outflow region too, indicating the
presence of shocked and hot gas in the outflow
\citep{Do02,gry99,bea10}. Different masers including H$_2$O, OH,
SiO and CH$_3$OH were identified or imaged in the Orion KL outflow
region, and the masers may be excited by shocks \citep [and the
references therein]{mat10,hir12}.
The properties of the molecular outflow are still under
investigation.
Recently,
\cite{zap12} observed an outflow in the Orion KL
region using SiO (J=8-7, v=0) line with the Atacama Large
Millimeter/Submillimeter Array (ALMA). The high velocity gas lobes
are shaped like a butterfly.
The driving source of the outflow in the KL region remains
uncertain. It may be produced by the disruption of massive stellar
system occurred $\sim$ 500 years ago, which ejected massive
stellar objects BN, I and n \citep{zap09,bally11}. While
\citet{bea08} suggested that the submm core SMA1 may host the
driving source of the outflow. The Source I was identified as the
driving source of the SiO (J=8-7, v=0) outflow \citep{zap12}. The
origin of the outflow needs to be further tested and explored.


Material infall toward cores or forming stars is another kind of
essential motion in star formation regions. Although the infall
studies progress more slowly than those of molecular outflows,
gravitational collapse candidates were frequently found toward
massive star formation regions with blue profile \citep [and the
references therein]{wu03,wu07,ch10}. Stronger collapse signature,
inverse P Cygni profile was detected early with high resolution
observations \citep [and the references
therein]{wel87,zha97,qin08,liu13}. The dynamical collapse or
inflow motion in this active star formation region were not
reported before.

To examine the dynamical processes and especially to uncover the
origin of the outflow in this region, observations with tracers of
high excitation temperature and critical density are needed. This
paper presents the results of an observational study of Orion KL
at the 1.3 mm with the ALMA. Using continuum emission and spectral
lines of CH$_{3}$CN (12$_{4}$-11$_{4}$)  and CH$_{3}$OH (8$_{-1,
8}$-7$_{0, 7}$) in both emission and absorption, an outflow with
multiple lobes in the KL region was detected. Inflow motion was
identified toward Source I and the hot core.

\section{Observations}
The data used in this paper was obtained from ALMA science
verification (SV) at band 6. The observations were made with
16$\times$12 m antennas of ALMA on January 20, 2012, with
baselines ranging from 18 to 253 k$\lambda$. The phase center was
at R.A. = $05^{h}35^{m}14^{s}.35$ and Dec. =
-$05\arcdeg22\arcmin35\arcsec.0$. There are 20 spectral windows
(SPW) of 1.875 GHz wide. Each spectral window consists of 3840
channels with a spectral resolution of 0.488 MHz ($\sim$ 0.7
km~s$^{-1}$). Callisto was used for bandpass calibration and flux
calibration while quasar J0607-085 provided the phase calibration.
The primary beam size of FWHM is about 30$\arcsec$ and the
emission of the KL region falls well inside of the FWHM.

The synthesized images with the calibrated data were carried out
by use of the Common Astronomy Software Applications (CASA)
package and MIRIAD \citep{sau95}. The continuum map was
constructed from the line-free channels of the five spectral
windows at highest frequency bands. The natural-weighted
beam size and the rms noise level for the continuum are
1$\arcsec.66\times1\arcsec.35$ with P.A.=-20$\arcdeg$.7 and 8 mJy
~beam$^{-1}$ (0.07 K), respectively. The CH$_{3}$CN
(12$_{4}$-11$_{4}$) and CH$_{3}$OH (8$_{-1, 8}$-7$_{0, 7}$) lines
were extracted from spectral windows "spw1" and "spw19",
respectively, while the $^{13}$CH$_{3}$CN (12$_{3}$-11$_{3}$) and
(13$_{3}$-12$_{3}$) transitions were identified in spectral
windows "spw5" and
 "spw6", respectively. bThe typical channel rms for lines is
 about 20 mJy ~beam$^{-1}$ per channel (0.18 K).

\section{Results}
\subsection{Continuum}
The continuum image of the 1.3 mm emission is presented in Figure
1a. Seven millimeter cores are identified and denoted with MM
followed by a number. MM1 is the strongest and largest one, which
corresponds to the hot core \citep{wil00}. The
others are located in the compact ridge. MM7 is far away from the phase center and is
excluded from the further analysis.
The known infrared sources were plotted on the map and shown by
crosses \citep{shu04}. Two dimensional gaussian fits were made
toward the continuum cores. The peak position is listed in column
2 of Table 1. Column 3 is the core size, here 414 pc was adopted
as the distance \citep{men07}.
The peak emission intensity and the total flux of the cores are
given in Columns 4 and 5 of Table 1.

$^{13}$CH$_{3}$CN (12$_{3}$-11$_{3}$) and (13$_{3}$-12$_{3}$)
transitions were extracted from each of the cores. The systematic
velocities (V$_{lsr}$), and the mean line widths ($\Delta$~V) of
the $^{13}$CH$_{3}$CN lines by Gaussian fitting are given in
columns 6 and 7 of Table 1. We also calculated the rotational
temperatures of the cores with the rotational temperature diagram
(RTD) method \citep{ara05,gold99,liu02,qin10} . Only the lines not
blended and with high S/N levels ($>5\sigma$) are used for RTD
analysis. Assuming that the local thermodynamic equilibrium (LTE)
holds and the lines are optical thin, T$_{rot}$ values are
obtained, which are listed in the eighth column of Table 1. Figure
1b shows the rotational temperature diagram of MM6 as an example.
bWe estimated the virial mass of the MM cores with the
relation \citep{mac88}:
$M_{vir}=210(R/pc)(\Delta V/kms^{-1})^{2}(M_{\odot})$
which is given in the last column of Table 1.
\subsection{Line profiles and spatial distribution of gas emissions}
Figure 2 shows the spectral lines of CH$_{3}$CN
(12$_{4}$-11$_{4}$) and CH$_{3}$OH (8$_{-1, 8}$-7$_{0, 7}$) from
the peak position of each dust core observed with ALMA. There are
two kinds of line profiles from these cores. One has the emission
distributed in both sides of the systemic velocity, e.g., the
lines at MM1 and MM2 positions. Both CH$_{3}$CN and CH$_{3}$OH
lines at MM1 have double peaks with the blue one stronger than the
red one, and the absorption dip is red shifted with respect to the
systemic velocity. These absorption dips are unlikely to be
caused by missing flux since the emission of the KL region falls well
inside the primary beam, especially the CH$_{3}$CN
(12$_{4}$-11$_{4}$) and CH$_{3}$OH (8$_{-1, 8}$-7$_{0, 7}$) lines
that we used have high excitation densities. Additionally both
lines from the core MM2 are symmetric, indicating that the
instrument effect cannot account for the absorption, which can be
interpreted as gas inflow motion toward the core (See more in
Sect. 3.4).


The other kind of line profiles show that most of gas emission is
located in one side of the systemic velocity. The spectral line at
MM4 has stronger emission on the blue side, and on the red side
the intensity decreases sharply from the peak. In contrast to the
MM4, the spectral emission of MM3, MM5 and MM6 are mainly on the
red side. These profiles show characteristics of high velocity gas
moving toward us or away from us, which are similar to the fast
moving molecular bullets near YSOs as shown in Figure 7 and Figure
9 of \citet {bach90}, but there are apparent differences. The high
velocity gas observed here are not molecular bullets ejected from
the immediate vicinity of the YSOs, but are jet propelled cores
(hereafter JPCs). Firstly, the JPCs in Orion KL show
strong dust emission with total flux ranging from 0.43 to 1.27 Jy,
which are very rare among the bullets in bipolar molecular
outflows or HH objects. Their virialization masses range between
those of the hot core and core MM2. While the total masses of
bullets near low-and high-mass YSOs are $\leq$ 0.07 M$_{\odot}$ in L1448
\citep{bach90,rich92,hat99}. Secondly, the molecular bullets near
YSOs are always highly collimated, and the blue and red bullets in
high or low mass star formation regions are located along an
outflow axis \citep{rich92,hat99}. In the Orion KL region, the
JPCs have no such bipolar structure, but appear to distribute
within the inner edge of the gas region. Thirdly, the JPCs are hot
with T$_{rot}$ of 100 to 182 K. Usually line profiles of bullets
in bipolar molecular outflows have been observed in low transition
lines of CO J=1-0, 2-1 \citep{bach90,hat99}. But for the JPCs in
Orion KL, no such profile was found in lines of CO J=1-0, 3-2,
HCO$^+$ J=1-0 and SO (3$_2$-2$_1$) \citep{eric82,olo82,Fri84}. In
addition, the spectral profiles of the JPCs are different from
those of molecular bullets in bipolar outflows near YSOs. The
former have single component while the later are usually
accompanied by a line from the host molecular core
\citep{bach90,rich92,hat99}, indicating that the JPCs in the Orion
KL region were not ejected by a YSO.
Their systemic velocity is not separated from the high velocity
emission, meaning that the original gas in the KL region was
impacted and propelled by the explosive jets
\citep{tay84,bally11,zap11}. These cores are not explosive jets
but jet-propelled cores.



The spectral features can also be seen from channel maps of
CH$_{3}$CN (12$_{4}$-11$_{4}$) line illuminated in Figure 3, where
the white contours denote the emission intensity averaging over 3
km~s$^{-1}$ velocity range. The gas emission at velocity $\le$
$\sim$(-13) km~s$^{-1}$ is contaminated by CH$_{3}$CH$_2$CN
(25$_{2,24}$-24$_{ 2,23}$) line.
The major bulk of the gas distribute within velocity range of -5
to 16 km~s$^{-1}$. A comparison with Figure 1(a) shows that as the
velocity increases, the emission first arises from the west of
MM1, and then from MM4 and south-east of MM5. From 1 to 16
km~s$^{-1}$, the morphology of the emission region looks like a
semi-wedge ring.  A complete wedge ring appears with velocity
ranging from 6 to 11 km~s$^{-1}$, which is similar to an angle
open to the north. These results show that the molecular gas
emissions of the KL region are physically related to different
velocity components including molecular JPCs and cores.

In the blue-shifted high velocity range of -14
to -2
km~s$^{-1}$, gas emission arises from MM4, the south-east of MM5
and the west of MM1, then extends to north-east and south. The
red-shifted high velocity gas emissions arise from MM6-MM5, the
south-west of MM1 and near MM2/MM3 with velocity range of $\sim$
19 to 22 km~s$^{-1}$. The components of the high velocity gas also
consist of emissions of JPCs and cores.


\subsection{High velocity gas -- an outflow with multiple lobes}
From Figure 2, the blue and red-shifted high velocity gas range
from -13 to -3 km~s$^{-1}$ and 15 to 24 km~s$^{-1}$, respectively.
The contours of the velocity integration intensities of the high
velocity gas measured with the CH$_{3}$CN (12$_{4}$-11$_{4}$) line
are presented in Figure 4, in which the crosses represent the
infrared sources.

The CH$_{3}$CN (12$_{4}$-11$_{4}$) outflow has two blue lobes and
five red lobes. The blue shifted high velocity gas peaks near MM4
and the west of MM1. The red shifted high velocity gas reaches the
maximum near MM5. The spectra of MM1 and Source I have core
emission profile with blue and red high velocity wings, but the
lines of the JPCs have much more obvious blue or red shifted
emissions (Figure 2 and Figure 5, see also next section). The
distribution of the high velocity gas suggests that the outflow
mainly overlaps or connects with JPCs. The dissimilarities between
the spectra of the cores and the JPCs also demonstrate that the
molecular outflow is not
 from MM1 or Source I.

The red outflow has 5 lobes and looks like a projection of
a bracelet, while the blue outflow has two lobes. The outflow
covers JPCs MM4-MM6. The south-west part of MM1 overlaps with the
outflow, and Source I is located outside of the outflow. Both MM1
and Source I can not be the origin or the driving source of the
outflow. The other YSOs in this region are located in different
lobes and can not serve as the driving source of the outflow too
(see Figure 4). These results demonstrate that the high velocity
gas in the KL region is not caused by a usual bipolar molecular
outflow derived by a YSO, but is mainly from distinct molecular
cores propelled by bullets from a burst. That is why blue and red
lobes cannot be separated, and the bipolar outflow detected in the
KL region with CO at the early period has poor collimation
\citep{kwan76,wu04}.

\subsection{Gravitational collapse toward Source I and inflow motion to the MM1}

To examine the line emissions of MM1 (the hot core) and Source I,
we plot the CH$_{3}$OH (8$_{-1, 8}$-7$_{0, 7}$) and
$^{13}$CH$_{3}$CN (13$_{3}$-12$_{3}$ lines at the boundaries of
MM1 and source I in Figure 5. The CH$_{3}$OH spectral profile from
Source I shows blue-shifted emission and red-shifted absorption
with respect to the systematic velocity of 5.5 km~s$^{-1}$. It
presents an inverse P Cygni profile, indicating that both the
gases behind and in front of the continuum source are moving
toward the continuum source.

The spectrum of CH$_{3}$OH at  MM1 has gas self absorption dip
which is red-shifted relative to the systematic velocity. The blue
peak of CH$_{3}$OH  line is stronger than the red one with a ratio
of T$_B$/T$_R$=2.16. For the line profile, the asymmetry parameter
$\delta V$ ($=V_{thick}-V_{thin})/\delta V_{thin}$) defined by
\citet{mar97} was calculated, giving $\delta V$  $\le$ -0.25 for
the blue profile, where V$_{thick}$ is the peak velocity of the
optical thick line, V$_{thin}$ and $\delta V_{thin}$ are the
systemic velocity and the line width of the optically thin line,
respectively. For the line pairs of the center of MM1, $\delta
V=(V_{thick}-V_{thin})/\delta V_{thin}$=-0.26 was obtained.  The
two parameters of T$_B$/T$_R$ and $\delta$ both satisfy the
criteria of the blue profile.

The optical thick and thin spectral lines from position (d), east
of MM1 have similar intensity and velocity partner to those from
the center of MM1. The ratio of T$_B$/T$_R$=2.13 and the
$\delta$V=-0.25 satisfy the criteria of a blue profile. The
signature is slightly weaker than that of the center one,
suggesting that inflow motion in MM1 is moving toward the core
center.

The line of the position (e), from MM1, has a profile that
belongs to those of JPCs and has blue-shifted high velocity gas. This
suggests that the west side of the core MM1 was
impacted by the explosive jets. The profiles of the lines from
north and south of MM1 seem to be affected by the explosive
outflow but not as significant as the one at position (e).

\section{Discussion}

\subsection{The JPCs}

As stated before, the molecular gas of the KL region actually
consists of different velocity components. Beside the cores, there
are JPCs mainly distributed in the compact ridge. There
are two kinds of line profiles produced from JPCs. One has blue
wings and the other has red wings. The high velocity wings in the
two kinds spectra are from -13 to -3 km~s$^{-1}$ and 15 to 24
km~s$^{-1}$ respectively, which are similar to those from SiO
(8-7, v=0) \citep{zap12}. Such high velocities may be excited by
C-type shock since they are below the critical velocities of
J-type shocks, which range from 25 to 45 km~s$^{-1}$. J-type
shocks are also with emissions primarily in the UV and optical,
while C-type shocks mainly excite the vibrational and high
rotational levels of molecules in infrared and radio emissions
\citep{wan92}. Unlike the spectra of molecular bullets near YSOs,
the emissions of ambient and high velocity gas of JPCs are not
separated. From Table 1 and Figure 5, one can see that the
V$_{lsr}$ of the JPCs are different. The spectra of JPCs with blue
wings have smaller V$_{lsr}$ while those with red wings have
larger V$_{lsr}$. If we take the average value of the peak
velocities of all the JPCs and cores in the KL region, 7.7
km~s$^{-1}$ as the systemic velocity of the KL region gas, then
the shifts of the peak velocities of the JPCs range from -2.4 to
2.3 km~s$^{-1}$. The JPCs are moving away or towards us at slow
speed.
The slow systemic motions of the JPCs may be caused by a rocket
effect when the JPCs were propelled by the jets.

\subsection{Explosive molecular outflow}
\subsubsection{High velocity molecular outflow driven by explosive jets}
The high velocity molecular outflow has multiple lobes (see Figure
4). The outflow gas mainly distributes in the ridge and only a
small part overlaps with the south-east of MM1 and the north of
MM2, indicating that the high velocity gas comes mainly from the
JPCs.
From Figure 2 and Figure 5 one can see that the blue shifted JPCs
are located east of the ridge and the red ones distribute more
widely. Unlike bipolar molecular outflows in which usually a YSO
driving the high velocity gas emission, every blue or red lobe in
the KL region harbors YSOs. Source I and IRn are located far from
the outflow center.
These results suggest that the outflow is unlikely to be driven by
some young stellar objects in the outflow region.

To examine the origin of the high velocity gas and shock effects on
the JPCs, we investigated the distribution of the IRAC [4.5/3.6] flux
ratio in the Orion KL region. The 4.5 $\mu m$ band emissions
detected with IRAC of Spitzer contain lines excited by shocks
\citep{cyg09}. A ratio
$>$ 1.5 represents shocks or jets \citep{ta10}. The ratio map with
color scale overlaid with the continuum map is displayed in Figure
6. Four infrared bubbles (IRBs) were identified, and the largest
and brightest one IRB1 is centered at BN and the emissions are
saturated at the inner region.
IRB2 is the second largest and is connected with IRB1 and IRB3.
IRB4 is located south of the region. Among the ALMA detected dust
cores, MM6 and MM5 are located at the outer part of IRB1 and IRB2
respectively. MM4 is located at the outer part of IRB 3. The
north-east of MM3 is connected with IRB4. These bubbles have the
flux ratio between $\lesssim$1.0 and 1.4, and seem to be swept out
by wind and jets. The JPCs are located outside the bubbles or
partially overlap with bubbles. It is notable that the strongest
and largest shocked area in the KL region is a large wedge ring,
and surrounds the KL gas wedge. This ring is well associated with
the H$_2$ 2.12 and 2.15 $\micron$ clumps \citep{sto98}, suggesting
shocked effects in this region.

To further probe the excitation mechanism of the molecular
outflow, we also plot the JPCs and cores of the KL region overlaid
with the 2.12 $\mu m$ H$_2$ jets \citep[and the references there
in]{bally11} in Figure 7. Clearly the JPCs do not overlap with the
finger-like jets. The
western side of the core MM1 where the shocked gas
(Figure 5) was detected partially overlaps with a finger-like jet.
These results indicate that the jets break into the region and the
cores are propelled or impacted. The gas is accelerated and
entrained, resulting in the multi-lobe molecular outflow.


We identified the BN object as the explosive center in the next
section. Here we estimate the dynamical age of the expanding
bubble. Using the distance from BN to the project center of the
high velocity gas as the average length of the jets
(8$^{\prime\prime}$.9) and using the average velocity of the
outflow, the dynamic time is estimated to be 1280 years, which is
consistent with the numerical experiments of gradational slingshot
\citep{chat12}.


\subsubsection{The explosive center}

 One astonishing characteristic seen from Figure 6 and Figure 7
is the distribution of YSOs. Almost all the YSOs in the ridge and
BN region are located at the edges or outside of the jets, while
all the YSOs are located within the bubbles or on the rings. BN is
located at the center of the largest bubble IRB1. IRc15, IRc16W,
IRc16S, IRc16N, IRc17, IRc6E, IRc6 and IRc6S are located in IRB1
\citep[also see][Fig.1]{shu04}. IRc3N and IRc3S are embedded in
the second largest bubble IRB2. IRc2, IRc7 and IRcn are located in
IRB3. IRc4 is located in IRB4.
Recent evidence of triggered star formation were found in a number
of bubbles \citep{wat08,deh03}. The ionizing stars responsible for
the bubbles are located in the wind-evacuated cavity. A neutral
material layer forms between the ionization front and the shock
front when the HII region expands. This layer may collapse,
fragment, and then form stars \citep{el77,chu07,wat08,deh03}. The
shocked layers can be produced by various mechanisms including
colliding clouds, expanding HII regions, stellar wind bubbles and
supernova remnants. For Orion BN/KL region, there are explosive
jets while around the BN has a hypercompact HII region
\citep{pla13}. The facts about the gas distribution, and the
position of BN in the bubble suggest that BN is very likely the
explosive or expanding center.

BN is the hottest and evolved one among the stellar sources in the
region. It is a B type star with emission variation
\citep{beck67}.
At the wavelengths 12.5 $\mu$m and shorter infrared wavelengths,
BN is the dominating object \citep [and the references
therein]{shu04}.


Besides the BN, there are many YSOs in the hot core and
the ridge. Theses YSOs were identified from near or middle
infrared observations around BN at different times and with
different sensitivities. Their numbering generally has no order.
Although the first 5 YSOs discovered were numbered with IR1-IR5
located south-east of BN (IR1) \citep{rie73}, the sources found
later can not be numbered following this way. For example, IR6 and
IR7 were detected, and surrounded by the BN, IR2 and IR3
\citep{Dow81}. Because of the large extinction at near and middle
infrared wavelengths, their positions may be not so accurate.
However, the following analysis showed that the evolutionary
states of these YSOs are correlated with their distance from the
BN object.

The YSOs of IRc6, IRc3 and IRc 20 are closest to BN, and are
likely more evolved than those in other places of the region.
Their mid-infrared emission are rather weak \citep{shu04} and
strong emissions at longer wavelengths were not seen.
At the northwest fringe of the wedge ring, there
is no cold gas emission detected with the inversion lines of
$^{15}$NH$_{3}$ (3,3) \citep{wil00}, meaning that IRc6 is really
hot.

The objects IRc2, IRn and IRc7 are located farther away from BN and
younger than those of IRc6, IRc3 and IRc20. IRc2 A+B, IRc7 and IRcn
have much stronger mid-infrared emissions than those of IRc6, IRc3
and IRc20 in the IRB2. VLA observations showed that a shell of water
masers was located at IRc2 A, the connected region of the infrared
cluster IRc2 and MM1 \citep{pla88}.

IRc4 is the farthest one from BN among the YSOs in the ridge. Infrared
spectroscopic observations show that IRc4 is separated from IRc2
\citep{ait81}.
It has little emission at middle infrared but has strong emission at
20 $\mu$m wavelength \citep[and the references therein]{shu04}.
More recently, SOFIA observations show that IRC4 has the strongest
emission at 37.1 $\mu m$ \citep{de12}, suggesting it is the
youngest one among the YSOs in this region.

The physical state of the gas in Orion-KL region show that BN is
the power source. Figure 3 shows that the cavity of the hot gas
wedge ring detected with CH$_{3}$CN (12$_{4}$-11$_{4}$) is open
toward BN object, sighting the powered direction. From Figure 6
one can see that the largest shock ring clearly opens to the BN
object. MM6, MM5 and MM3 are located along the symmetric axis of
the wedge. Such structure can be seen in images of other
wavelengths. Infrared continuum measurements at 2-30 $\mu$m with
resolution 2$\arcsec$---2$\arcsec$.2 show that the KL region is
like a clumpy cavity with a diameter about 10$\arcsec$
\citep{Wynn84}. It is also shaped like a wedge ring. The most of
north infrared emission is located in its west fringe and covers
IRc6. However in this region there is no emission of the rather
cold gas detected with the inversion line of $^{15}$NH$_{3}$
(3,3), meaning the gas closest to the BN is hot.

The evolutionary states of the YSOs in the ridge correlate with
their distances from the BN object, that also support that the BN
is the powering source of the KL region.

For the explosive center of the Orion BN/KL region, various
mechanisms were suggested (see Sect. 1). In this work we report
some observational characteristics for BN as the explosive center.
More theoretical analysis and observation are needed to confirm
our identification.

\subsection{Dynamics of Source I and the hot core MM1}
The inverse P Cygni profile of line (b) shown in Figure 5
indicates gravitational collapse toward Source I. The material
infall velocity can be estimated as 1.5 km~s$^{-1}$ from the
V$_{sys}$ and V$_{obs}$ of the absorption peak \citep{wel87}. Here
we take the V$_{sys}$ of Source I as same as that of the core MM1.
Assuming the inflow is free fall \citep{pi14}, the mass
accretion rate of 1.2$\times$10$^{-3}$M$_{\sun}$/Yr can be
obtained with
  dM/dt=3V$_{in}^3$/2G,
which is much lager than those of low mass cores \citep{mye96}.
The result is consistent with the mass of Source I, which is about
10 M$_{\sun}$ \citep{tes10,liu13}.

For the core MM1, the lines at the center and position (d) have blue
profiles showing inflow motion. The infall velocity is calculated
with the model of \citet{mye96}:
\begin{equation}
V_{in}\approx\frac{\sigma^{2}}{v_{red}-v_{blue}}\textrm{ln}\left(\frac{1+eT_{BD}/T_{D}}{1+eT_{RD}/T_{D}}\right)
\end{equation}
where $T_{D}$ is the brightness temperature of the dip (assumed it
is optically thick), $T_{BD}$ and $T_{RD}$ are the height of the
blue and red peaks above the dip, respectively. The velocity
dispersion $\sigma\approx2.7$ km~s$^{-1}$ is obtained from the
optically thin $^{13}$CH$_{3}$CN(13$_{3}$-12$_{3}$) line. Thus the
infall velocities inferred from CH$_{3}$CN (12$_{4}$-11$_{4}$) and
 CH$_{3}$OH (8$_{-1, 8}$-7$_{0, 7}$) lines are $\sim$0.4 and $\sim$0.8 km~s$^{-1}$,
 respectively. Taking the average value $\sim$0.6 km~s$^{-1}$ as V$_{in}$, the mass accretion
 rate of  8.0$\times$10$^{-5}$M$_{\sun}$/Yr was obtained.  Both the infall
 velocity and the mass accretion rate are smaller than those of Source I
 but still larger than those of low mass cores \citep{mye96}.

Signatures of gravitational collapse or inflow motion were
detected in Orion KL region for the first time (See Sect.1).
For Source I, the material infall velocity and mass accretion rate
are computable with those of high mass sources \citep [see][and
the references therein]{liu13}. Early, an outflow from the Source
I was detected with v=0, J=2-1 line of SiO \citep{pla09}, which is
from NE to SW and is mainly distributed within the hot core.
Around Source I, an ionized disk was imaged with radio and 7 mm
continuum emission \citep{rei07}. Recently a hot and neutral
circumstellar disk around Source I was detected with ALMA
\citep{hir14}. The collapse signature, disk, and outflow suggest
that the high-mass YSO in Source I forms in similar way to those
of low mass stars.

Blue profiles were detected in the hot core, indicating that the
inflow motion is weaker than that in Source I. Previous
observations showed that inflows occur more frequently and more
strongly in UCH{\sc ii} regions than UCH{\sc ii} precursors \citep
[see][and the reference therein]{wu07}. The inflow difference of
Source I and MM1 is attributed to the fact that Source I is a
radio source while there is no YSO found yet in the hot core.

\section{Conclusion}
We have studied Orion KL region with 1.3 mm continuum and lines of
CH$_{3}$CN (12$_{4}$-11$_{4}$), $^{13}$CH$_{3}$CN
(12$_{3}$-11$_{3}$),(13$_{3}$-12$_{3}$) and CH$_{3}$OH (8$_{-1,
8}$-7$_{0, 7}$) observed with ALMA. Seven dust cores MM1 to MM7
were detected. Emissions of IRAC 3.6 and 4.5 $\mu$m was included
in the analysis. Our main findings are as the following:

1. Molecular emissions of CH$_{3}$CN (12$_{4}$-11$_{4}$) and
CH$_{3}$OH (8$_{-1, 8}$-7$_{0, 7}$) were detected from the dust
cores. The derived rotation temperatures T$_{rot}$ are from 119 to
198 K. Lines from MM1 and MM2 have usual profiles of cores, and
those from MM3, MM4, MM5 and MM6 show jet propelled line
characteristics. The line from the western area of MM1 also
presents the JPC profile similar to that of MM4. The gas emission
of the Orion KL region consists of these JPCs and the cores. The
morphology of the major body of the gas emission is similar to a
wedge ring.

2. The high velocity gas ranges from -13 to -3 km~s$^{-1}$ and 15
to 24 km~s$^{-1}$ for the blue and red wings respectively. The
morphology of the outflow shows multiple lobes.
The high velocity gas mainly comes from the JPCs and overlaps with
MM1 at its south-west and with the north of MM2. The
outflow does not seem to be driven by YSOs like usual bipolar outflows but
origins from an explosion.

3. Four infrared bubbles were found with the ratio of [4.5]/[3.6]
from infrared emissions with IRAC, Spitzer, which are related to
molecular bullets and contain the YSOs in the region. The
properties of the BN object, the possible formation mechanism and
evolutionary states of the YSOs as well as the distribution of
molecular gas in this region were analyzed. Results seem to
support that the BN object is the explosive center and the
explosion occurred 1300 yrs ago.

4. Properties and spectral profile of Source I and core MM1 are
presented. An inverse P Cygni profile was found toward Source I
showing gravitational collapse. The infall velocity and mass
infall rate are 1.5 km~s$^{-1}$ and
1.2$\times$10$^{-3}$M$_{\sun}$/Yr respectively. Inflow motion
shown in blue profile was detected toward MM1 (hot core), which
has infall velocity 0.6 km~s$^{-1}$ and mass infall rate
8.0$\times$10$^{-5}$M$_{\sun}$/Yr.

\section*{Acknowledgement}
\begin{acknowledgements}
We are grateful to the staff of ALMA for the observations. We
thank Jony Bally to send us the fits data of the H$_2$ jet image
and Hongping Du for the language improvement. We would like also
thank the anonymous referee for the constructive suggestions and
comments to improve the paper. This paper makes use of the
following ALMA data: ADS$\diagup$JAO.ALMA$\#$ 2011.0.00009.SV.
ALMA is a partnership of ESO (representing its member states), NSF
(USA) and NINS (Japan), together with NRC (Canada) and NSC and
ASIAA (Taiwan), in cooperation with the Republic of Chile. The
Joint ALMA Observatory is operated by ESO, AUI$\diagup$NRAO and
NAOJ. This work was supported by the China Ministry of Science and
Technology under State Key Development Program for Basic Research
(2012CB821800), the grants of NSFC number 11373009 and 11373026
and Midwest universities comprehensive strength promotion project
(XT412001, Yunnan university).
\end{acknowledgements}

\begin{figure}
\includegraphics[angle=90,scale=.50]{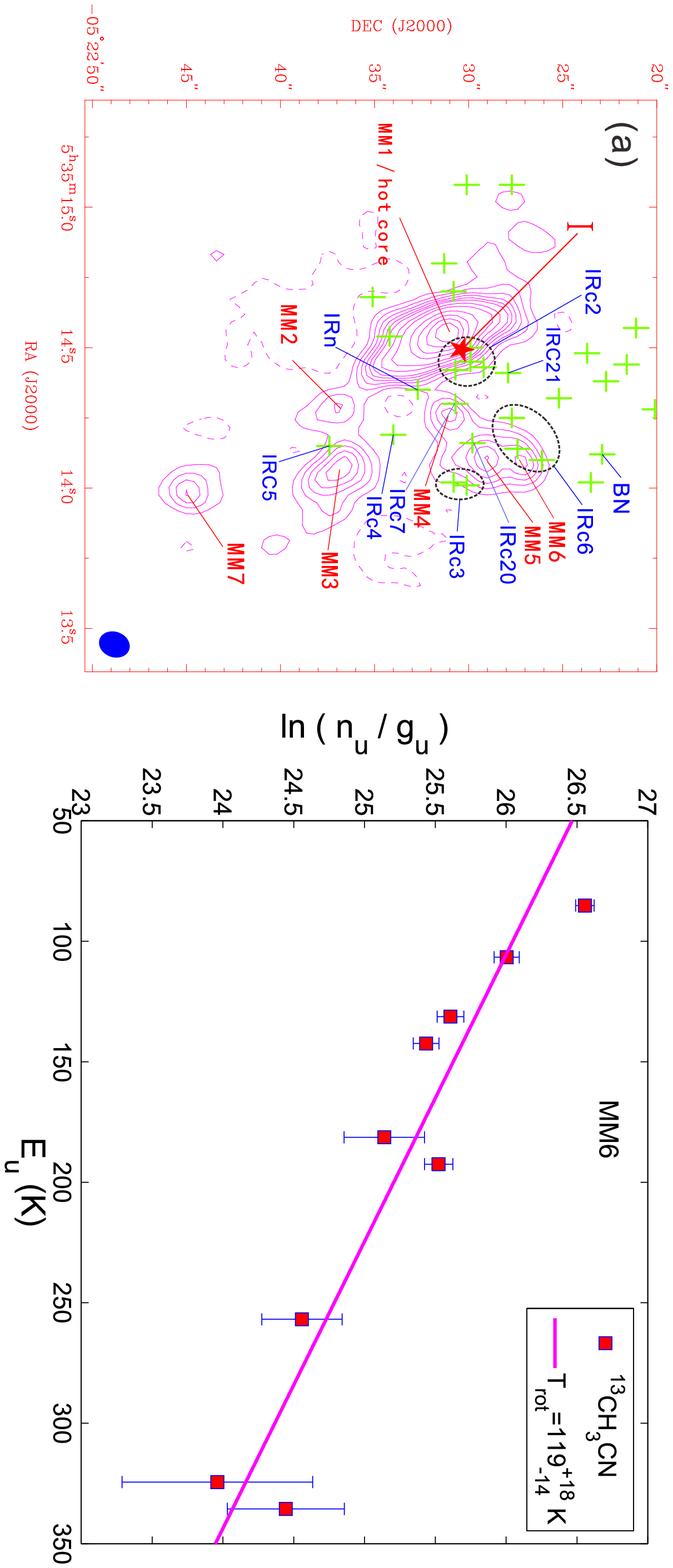}
\caption{(a). 1.3 mm continuum emission map of Orion KL. The contour
levels are (-5,5,10,15,20,25,30,40,50,70,90,110,130)$\times$0.01
Jy$\sim$beam$^{-1}$. The dust cores are labeled as MM1-7. The near
infrared sources \citep{shu04} are marked with "crosses". (b).
Rotational temperature diagram of $^{13}$CH$_{3}$CN at MM6.}
\end{figure}

\begin{figure}
\includegraphics[angle=-90,scale=.50]{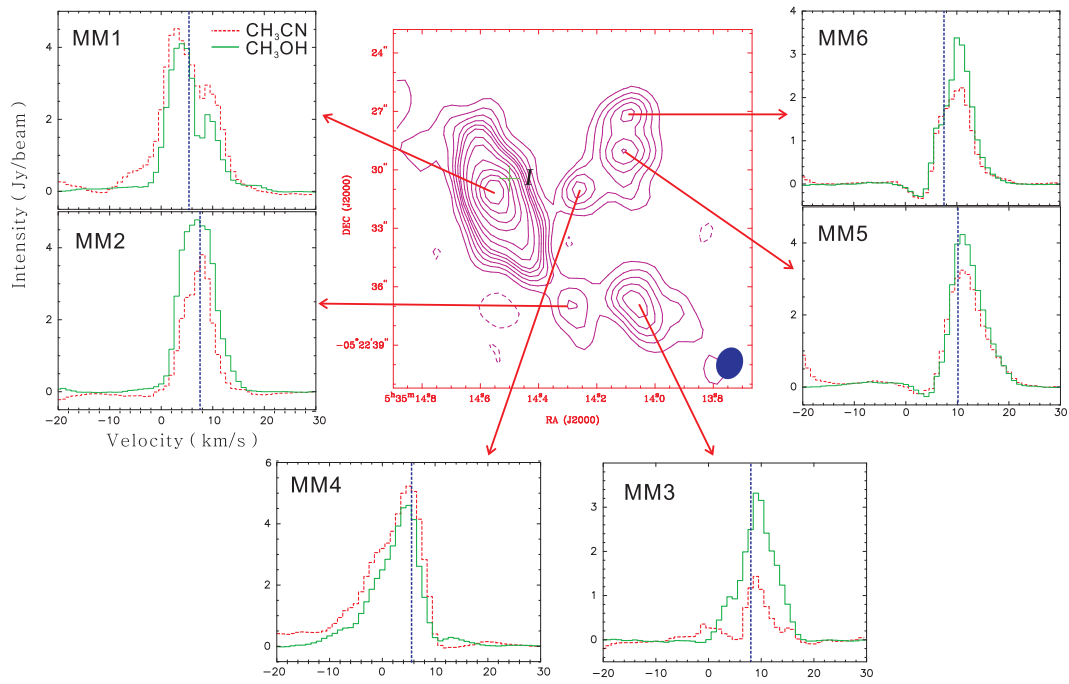}
\caption{Source averaged spectra at each dust core. The CH$_{3}$CN
(12$_{4}$-11$_{4}$) and CH$_{3}$OH (8$_{-1, 8}$-7$_{0, 7}$) spectra
are shown in red dashed lines and green solid lines, respectively.
The systemic velocities of each dust core are inferred from
$^{13}$CH$_{3}$CN (13$_{3}$-12$_{3}$) lines and marked with blue dashed
lines.}
\end{figure}

\begin{figure}
\includegraphics[angle=-90,scale=.75]{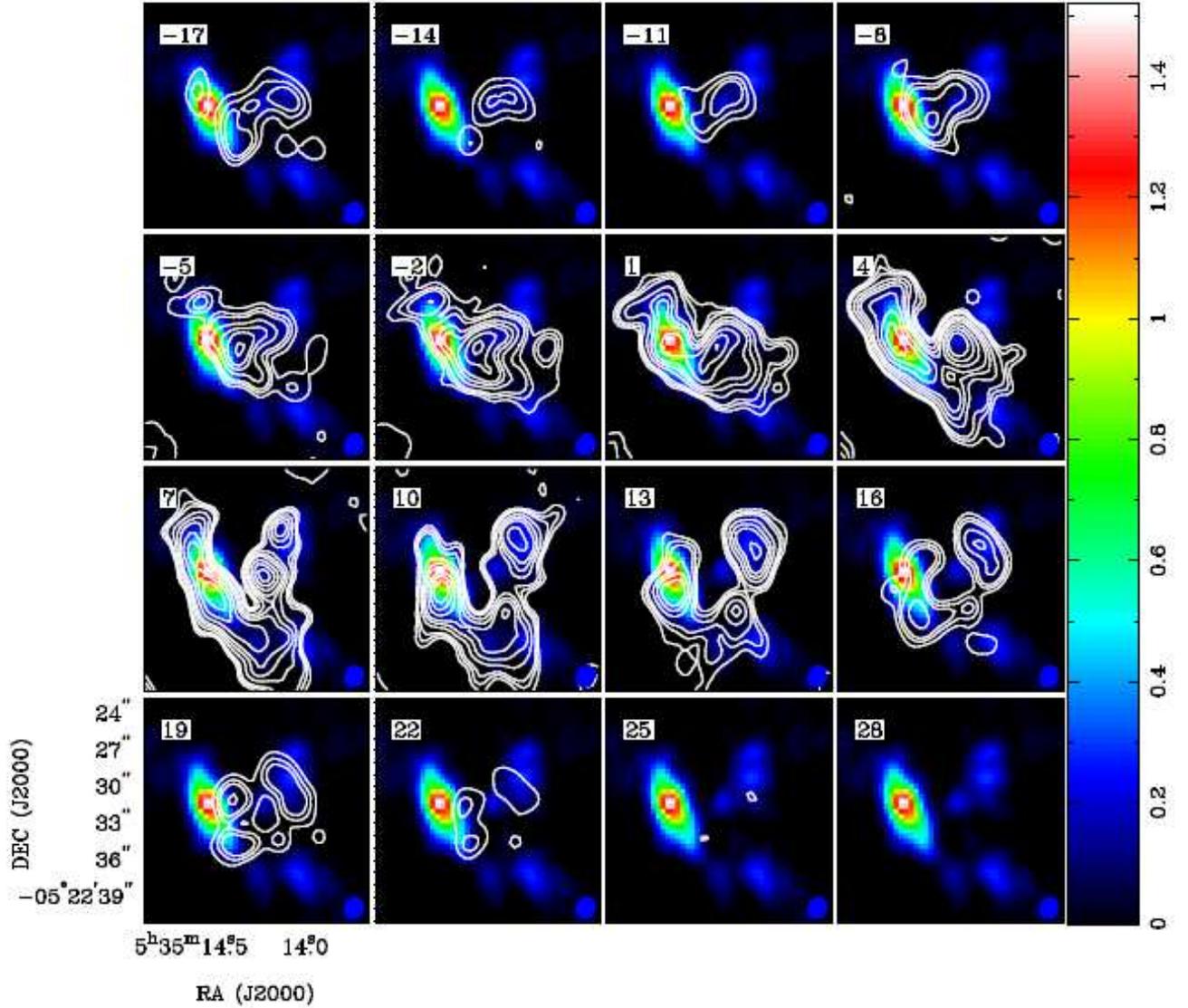}
\caption{Channel map of CH$_{3}$CN (12$_{4}$-11$_{4}$) overlaid on
the 1.3 mm continuum map. The channel wide is 3 km~s$^{-1}$. The
center velocity is labeled at the upper-left corner of each box. The
contour levels are (3,6,9,18,27,36,54,72,90,108,126) $\times$0.05
Jy~beam$^{-1}$.}
\end{figure}


\begin{figure}
\begin{minipage}[c]{0.5\textwidth}
  \centering
  \includegraphics[angle=-90,scale=.70]{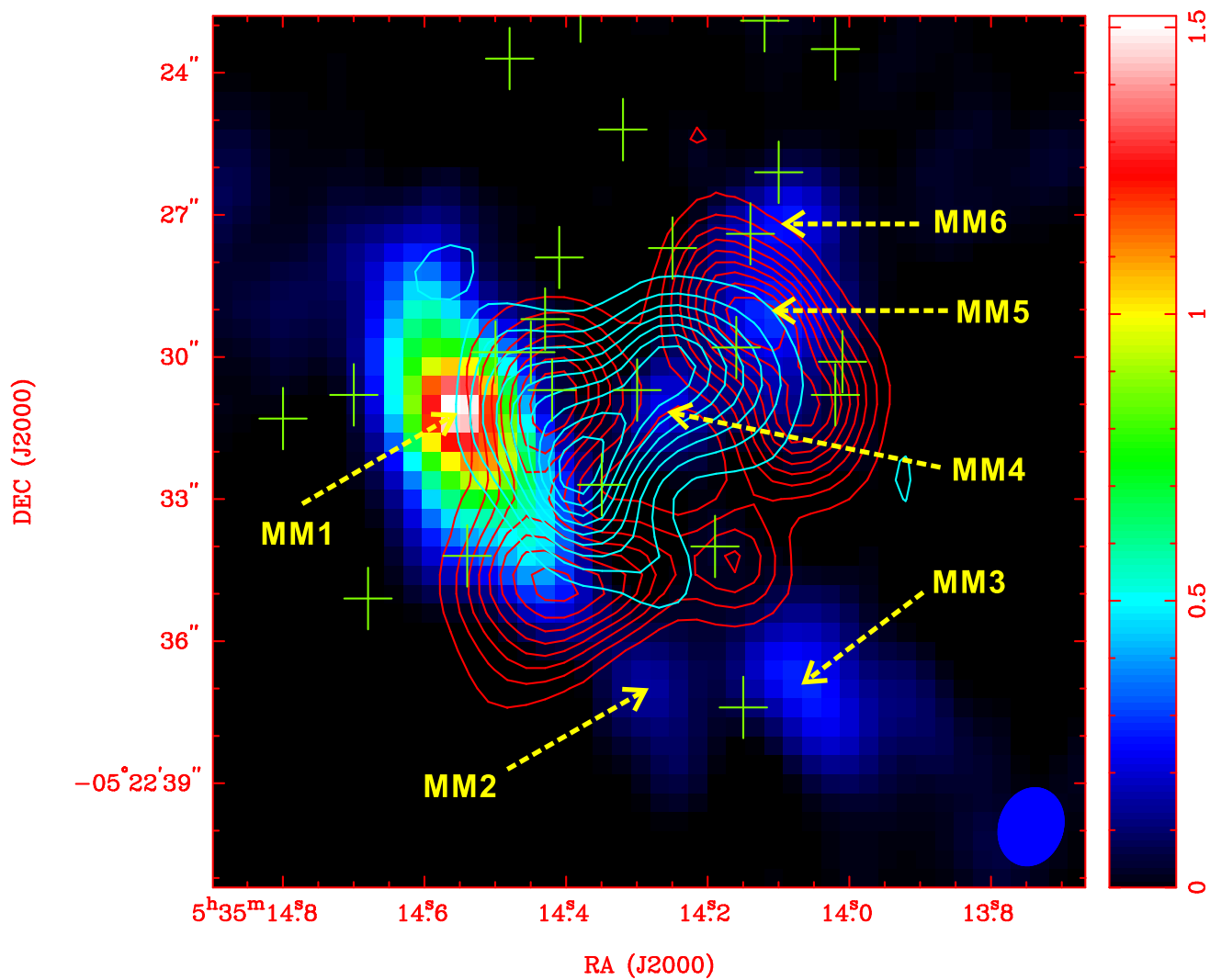}
\end{minipage}
\caption{Integrated intensity contours of high velocity CH$_{3}$CN
(12$_{4}$-11$_{4}$) emission. The background image is the 1.3 mm
continuum emission. The red contours represent redshifted gas,
which is integrated from 15 to 24 km~s$^{-1}$. The blue contours
show the blueshifted gas, which is integrated from -13 to -3
km~s$^{-1}$. The crosses indicate infrared sources.}
\end{figure}

\begin{figure}[!bht]
\includegraphics[angle=90,scale=.50]{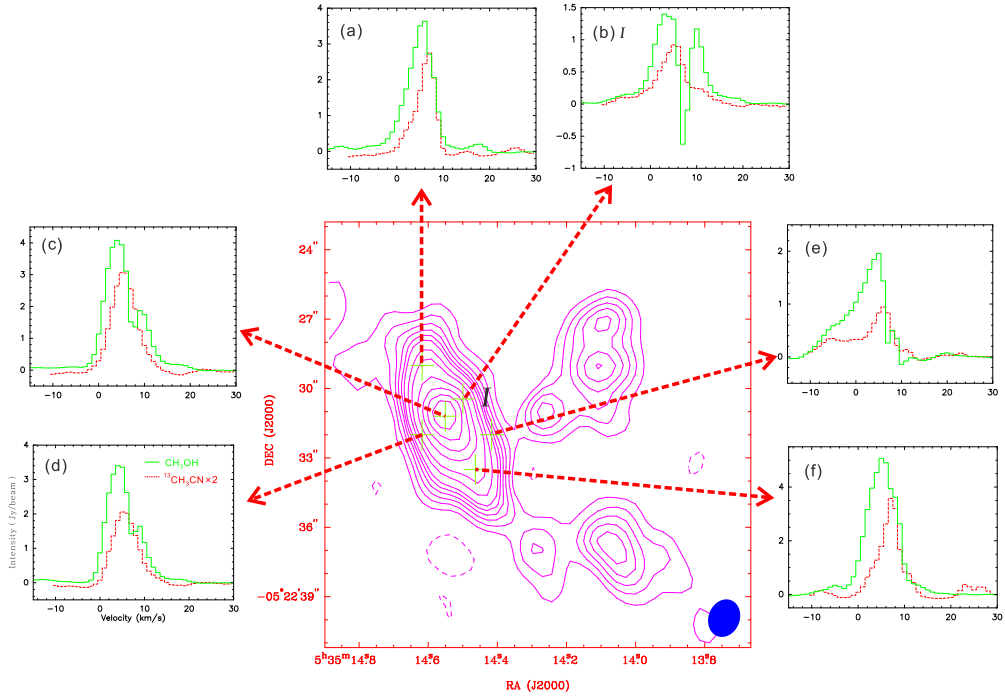}
\caption{Spectra of CH$_{3}$OH (8$_{-1, 8}$-7$_{0, 7}$)  and
$^{13}$CH$_{3}$CN (13$_{3}$-12$_{3}$) at different positions of MM1 in green and red colour respectively. The spectrum position is indicated at
the upper-left corner of each panel.}
\end{figure}


\begin{figure}
\includegraphics[angle=90,scale=.70]{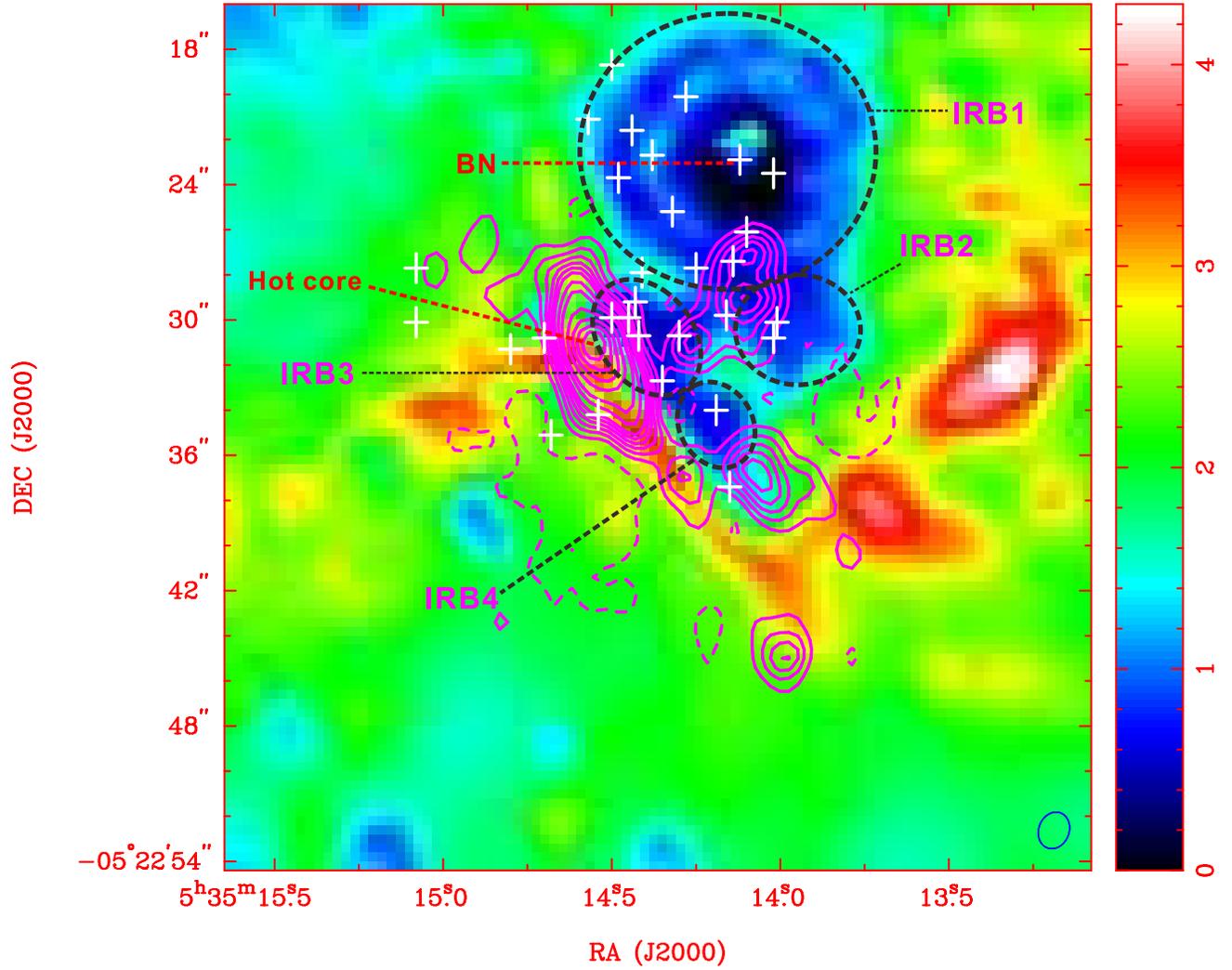}
\caption{[4.5/3.6] ratio image is shown in color. The four infrared
bubbles are depicted with ellipses and names as IRB1-4. The crosses
represent the near-infrared sources. The contours are 1.3 mm
continuum emission. }
\end{figure}

\begin{figure}
\includegraphics[angle=-90,scale=.70]{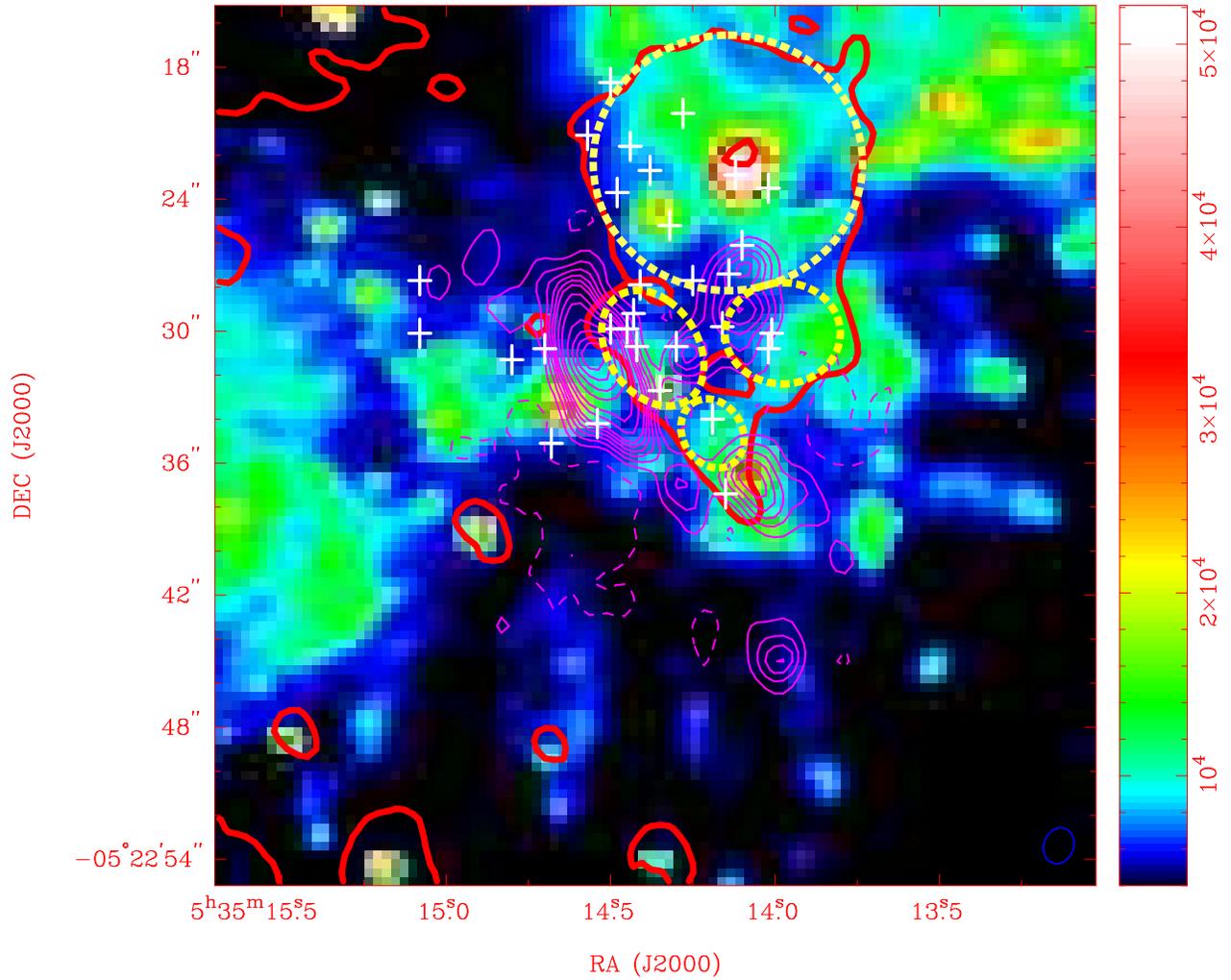}
\caption{The dust JPCs and cores of the Orion KL region (red
contours) and the finger-like jets (green images \citep[and the
references there in]{bally11}). The ellipses show the infrared
bubbles and the red line indicates the 4.5/3.6 ratio $>$ 1.5 (Also
see Figure 6).}
\end{figure}

\clearpage

\begin{deluxetable}{ccrrrrrrrrrrrrrrrcl}
\tabletypesize{\scriptsize} \tablecolumns{15} \tablewidth{0pc}
\tablecaption{Parameters of the continuum sources. } \tablehead{
 \colhead{Name} & \colhead{Offset}
&\colhead{R=$\sqrt{ab}$} &\colhead{I$_{peak}$} & S$_{\nu}$  &\colhead{V$_{lsr}$} &\colhead{$\Delta$~V} &\colhead{T$_{rot}$} &\colhead{M$_{virial}$}  \\
\colhead{}  & \colhead{($\arcsec,\arcsec$)} & \colhead{(10$^{-3}$
pc)}  &\colhead{(Jy~beam$^{-1}$)} & (Jy) & \colhead{(km~s$^{-1}$)}
& \colhead{(km~s$^{-1}$)} & \colhead{(K)} & \colhead{(M$_{\sun}$)}
} \startdata
MM1/hot core &  (2.90$\pm$0.05,3.61$\pm$0.08)   &  6.6   &   1.34(0.06) &  6.18  &     5.4(0.3) & 6.5(0.3) & 198$^{+36}_{-26}$    &      59   \\
MM2          &  (-1.07$\pm$0.11,-2.04$\pm$0.16) &    2.3   &  0.15(0.02) &  0.38  &     7.5(0.1) &3.1(0.1) & 156$^{+51}_{-31}$    &       5    \\
MM3          &  (-4.49$\pm$0.12,-1.97$\pm0.12$) &   3.8   &   0.28(0.03) &  1.02  &     8.0(0.2) &2.4(0.2) &  100?                &       5    \\
MM4          &  (-1.58$\pm0.14$,4.09$\pm$0.13)  &   2.4   &   0.22(0.03) &  0.43  &     5.5(0.2) &5.9(0.8) & 182$^{+81}_{-43}$    &      18   \\
MM5          &  (-3.61$\pm$0.07,6.09$\pm$0.15)  &     6.7   & 0.28(0.02) &  1.27  &    10.1(0.1) &5.4(0.2) & 157$^{+43}_{-28}$    &      41    \\
MM6          &  (-3.72$\pm$0.09,7.58$\pm$0.12)  &    2.2   & 0.28(0.02) &  0.61  &     7.5(0.3)  &5.7(0.3) & 119$^{+18}_{-14}$    &      15    \\
\enddata

\end{deluxetable}

\end{document}